\documentclass[aps,prx,twocolumn,10pt,footinbib,superscriptaddress,amsmath,amsfonts,amssymb,floatfix,notitlepage]{revtex4-2} 

\usepackage{graphicx}
\usepackage{dcolumn}
\usepackage{bm}
\usepackage{xcolor}
\usepackage{multirow}
\usepackage[percent]{overpic}
\usepackage[newcommand]{ragged2e}
\usepackage{rotating}
\usepackage{tikz}
\usepackage{gensymb}
\usepackage{mathtools}
\usetikzlibrary{arrows.meta}

\usepackage{comment}
\usepackage{url}

\usepackage{caption}
\DeclareCaptionJustification{justified}{\justifying}
\usepackage{subcaption}

\DeclareMathAlphabet\mathbfcal{OMS}{cmsy}{b}{n}

\usepackage{xr-hyper}
\usepackage{hyperref}
\makeatletter
\newcommand*{\addFileDependency}[1]{
  \typeout{(#1)}
  \@addtofilelist{#1}
  \IfFileExists{#1}{}{\typeout{No file #1.}}
}
\makeatother

\newcommand*{\myexternaldocument}[1]{%
    \externaldocument{#1}%
    \addFileDependency{#1.tex}%
    \addFileDependency{#1.aux}%
}
\myexternaldocument{Supplementary_Information.tex}
\myexternaldocument{Z_hyperref}

\title{Competing Gauge Fields and Entropically-Driven Spin Liquid to Spin Liquid Transition in non-Kramers Pyrochlores}

\begin{abstract}
Gauge theories are powerful theoretical physics tools that allow complex phenomena to be reduced to simple principles, and are used in both high-energy and condensed matter  physics. 
In the latter context, gauge theories are becoming increasingly popular for capturing the intricate spin correlations in spin liquids, exotic states of matter in which the dynamics of quantum spins never ceases, even at absolute zero temperature. 
We consider a spin system on a three-dimensional pyrochlore lattice where  emergent gauge fields not only describe the spin liquid behaviour at zero temperature but crucially determine the system's temperature evolution, with distinct gauge fields giving rise to different spin liquid phases in separate temperature regimes. 
 Focusing first on classical spins, in an intermediate temperature regime, the system shows an unusual coexistence of emergent vector and matrix gauge fields where the former is known from classical spin ice systems while the latter has been associated with fractonic quasiparticles, a peculiar type of excitation with restricted mobility. 
 Upon cooling, the system transitions into a low-temperature phase where an entropic selection mechanism depopulates the degrees of freedom associated with the matrix gauge field, rendering the system spin ice like. 
 We further provide numerical evidence that in the corresponding quantum model, a spin liquid with coexisting vector and matrix gauge fields has a finite window of stability in the parameter space of spin interactions down to zero temperature. 
 Finally, we discuss the relevance of our findings for non-Kramers magnetic pyrochlore materials.

\end{abstract}

\begin{document}
\preprint{APS/123-QED}

\title{Competing Gauge Fields and Entropically-Driven Spin Liquid to Spin Liquid Transition in non-Kramers Pyrochlores}

\author{Daniel Lozano-G\'omez}
\affiliation{Department of Physics and Astronomy, University of Waterloo, Waterloo, Ontario N2L 3G1, Canada}
\affiliation{Institut f\"ur Theoretische Physik and W\"urzburg-Dresden Cluster of Excellence ct.qmat}
\affiliation{Technische Universit\"at Dresden, 01062 Dresden, Germany}
\author{Vincent Noculak}
\affiliation{Dahlem Center for Complex Quantum Systems and Fachbereich Physik, Freie Universit\"at Berlin, 14195 Berlin, Germany}
\affiliation{Helmholtz-Zentrum Berlin f\"ur Materialien und Energie, Hahn-Meitner-Platz 1, 14109 Berlin, Germany}
 \author{Jaan Oitmaa}
 \affiliation{School of Physics, The University of New South Wales, Sydney 2052, Australia}
 \author{Rajiv R. P. Singh}
 \affiliation{ Department of Physics, University of California Davis, California 95616, USA}
\author{Yasir Iqbal}
\affiliation{Department of Physics and Quantum Centre of Excellence for Diamond and Emergent Materials (QuCenDiEM), Indian Institute of Technology Madras, Chennai 600036, India}
\author{Johannes Reuther}
\affiliation{Dahlem Center for Complex Quantum Systems and Fachbereich Physik, Freie Universit\"at Berlin, 14195 Berlin, Germany}
\affiliation{Helmholtz-Zentrum Berlin f\"ur Materialien und Energie, Hahn-Meitner-Platz 1, 14109 Berlin, Germany}
\affiliation{Department of Physics and Quantum Centre of Excellence for Diamond and Emergent Materials (QuCenDiEM), Indian Institute of Technology Madras, Chennai 600036, India}
\author{Michel J. P. Gingras}
\affiliation{Department of Physics and Astronomy, University of Waterloo, Waterloo, Ontario N2L 3G1, Canada}

\maketitle

Gauge symmetries and their embodiment within pertinent mathematical frameworks constitute a quintessential aspect of some of the most fundamental theories of physics, ranging from Maxwell's electromagnetism and Einstein's general relativity to the Standard Model of particle physics. In such fundamental theories, different types of gauge fields usually coexist, but are of physical relevance only within a characteristic energy scale. Over the past forty years or so, the application of gauge theories in condensed matter physics to describe strongly correlated electron and magnetic (spin) systems has steadily grown~\cite{Wegner-2003,Lee-2006,wenQuantumFieldTheory2004}. In the contemporary field of highly frustrated 
magnetism~\cite{Springer_frust, balentsSpinLiquidsFrustrated2010b}, gauge symmetries can emerge from the energetic constraints on the allowed spin orientations that are imposed by competing (i.e., frustrated) spin-spin interactions. The latter can significantly enhance the magnitude of thermal and quantum fluctuations, thus undermining the development of conventional long-range magnetic order, but still stabilizing strong nontrivial spatio-temporal correlations between the spins, producing a liquid-like state of sorts \textendash~\emph{a spin liquid}~\cite{
Springer_frust,balentsSpinLiquidsFrustrated2010b,zhouQuantumSpinLiquid2017,Gingras_2014,zhouQuantumSpinLiquid2017,broholmQuantumSpinLiquids2020,savaryQuantumSpinLiquids2016}.

Gauge theories have proven powerful schemes to describe a wide range of spin liquids~\cite{Misguich-2002,Benton-2021,wietek2023quantum,Song-2019}.
In particular, the usage of such theories has allowed to uncover various spin liquid states harboured by magnetic systems whose spins reside on the vertices of the three-dimensional pyrochlore lattice of corner-sharing tetrahedra and to expose their exotic properties~\cite{
Henley-2010,Castelnovo-2012,YanRank2U1PhysRevLett.124.127203,Benton2016Pinch-line-singularity}.
For example, the spin liquid state found at low temperatures in spin ice materials~\cite{Castelnovo-2012} (e.g. R$_2$M$_2$O$_7$ (R=Ho, Dy; M=Ti, Sn, Ge)~\cite{Zhou_pressure_spin_ice})
 is characterized by constrained orientations of the magnetic moments that are akin to an effective Gauss' law describable by an emergent gauge field~\cite{Henley-2010,Castelnovo-2012}.
 This elegant description has direct experimental consequences, namely, the spin-spin correlations show non-analytical ``pinch point'' singularities in reciprocal (momentum) space that are revealed in neutron scattering experiments ~\cite{Fennelscience.1177582,Morris_science}, thus bearing witness to the underlying gauge symmetry describing the spin ice state~\cite{ Henley_powerlaw_2005,ChungKristianPhysRevLett.128.107201}.

Classical spin liquids (CSLs) have extensive ground state degeneracy arising from fine-tuned sets of spin-spin interactions~\cite{Bergman-2007,Mizoguchi-2018,Benton-2021} and are potential harbingers of quantum spin liquids (QSLs) upon consideration of the spins' quantum dynamics~\cite{BuessenDiamond,Iqbal19,Niggemann-2023}.
However, in the classical limit, ground state degeneracies are generically lifted by arbitrary small (perturbative) symmetry-allowed interactions thence typically inducing long-range magnetic order. 
One can flip this perspective around to anticipate that CSLs, and thus QSLs, ought to often manifest themselves at the phase boundary between magnetically ordered classical ground states, as noted in prominent examples~\cite{Kolley2015,wietek2023quantum}.
Within this line of thought, and inspired by the successes of the gauge theory description of spin ice~\cite{Henley-2010,Castelnovo-2012}, a number of works on pyrochlore spin systems have shown that diverse gauge symmetries can arise at the boundary of competing classical long-range ordered phases, signalling new types of CSLs~\cite{YanRank2U1PhysRevLett.124.127203,Benton2016Pinch-line-singularity} and thus, potentially novel QSLs.
Examples include reports of a ``pinch-line spin liquid''~\cite{Benton2016Pinch-line-singularity} 
and a rank-2 U(1) spin liquid~\cite{YanRank2U1PhysRevLett.124.127203}, both of which described by emergent tensorial gauge fields.
The latter spin liquid is of particular interest because it is described by a symmetric rank-2 tensor gauge field akin to that present in  theories of fractons, spin excitations that can only propagate on subdimensional spaces~\cite{pretkoFractonPhasesMatter2020}.

The aforementioned successes beg the question: ``what new spin-liquid physics, at the classical or quantum level, may be evinced when \emph{three} classical phases meet at a triple point?''. This is the question we investigate in this paper by considering an effective spin-1/2 model for pyrochlore magnets of interacting non-Kramers rare-earth ions (i.e. that possess an even number of electrons)~\cite{Onoda_Kramers_PhysRevB.83.094411,
Lee_kramers_PhysRevB.86.104412,Rau-2019}
in a region of spin-spin couplings parameter space where three phases meet: one magnetic dipolar (spin ice) spin liquid phase and two electric quadrupolar long-range ordered 
phases~\cite{Onoda_Kramers_PhysRevB.83.094411,
Lee_kramers_PhysRevB.86.104412,Rau-2019}. We refer to this magnetic triple point as a dipolar-quadrupolar-quadrupolar (DQQ) point.
We find that the gauge symmetries at that point are enlarged due 
to competing and energetically degenerate 
rank-1 ($R_1$) and rank-2 ($R_2$) gauge fields, resulting in a novel spin liquid that we refer to as a $R_1$-$R_2$ spin liquid.
Most interestingly, we find that upon cooling, the rank-2 gauge field freezes out while the rank-1 U(1) spin ice liquid gets progressively entropically selected below a crossover temperature $T^*$.
We thus observe a novel phenomenon of a spin liquid to spin liquid transition in a magnetic system 
that is solely driven by temperature for a fixed spin Hamiltonian and not by tuning a parameter of the Hamiltonian as happens in the models considered in Refs.~\cite {yan2023experimentally,sanders2023vison}.Here, we refer to ``spin liquid'' as a cooperative paramagnetic phase~\cite{Villain} that does not spontaneously break any of the symmetries of the Hamiltonian.
Such thermodynamic behavior reminds one of the liquid-to-liquid transition observed in some atomic and molecular liquids~\cite{P-L2L,S-L2L}. 
We note that recent work~\cite{Hallen_Liquid_to_nematic_2024} found a temperature-driven transition separating two phases lacking long-range magnetic (i.e. dipolar) order.
However, in contrast with  Ref.~\cite{Hallen_Liquid_to_nematic_2024} 
wherein the two phases considered are a spin liquid phase and a symmetry-broken nematic phase, in the model under investigation in the present work, both phases preserve all symmetries of the parent Hamiltonian down to zero temperature.

Incorporating the effects of quantum spin fluctuations in our model, we find a window of spin-spin couplings close to the triple point where the system fails to display long-range magnetic order, thus providing evidence for a putative QSL.
These results are not only interesting from a strictly theoretical point of view, but may be of significant relevance for understanding the highly paradoxical non-Kramers Tb$_2$M$_2$O$_7$ (M=Ti, Sn, Ge) pyrochlore magnets~\cite{Rau-2019,Gingras_2014}.
This may be particularly so  for  Tb$_2$Ti$_2$O$_7$, which has defied  understanding since it was first studied~\cite{Gardner-Tb2Ti2O7}, and for which recent  work~\cite{TakatsuPRL2016TbtiO} proposed that this material may actually reside in the vicinity of such a DQQ triple point.

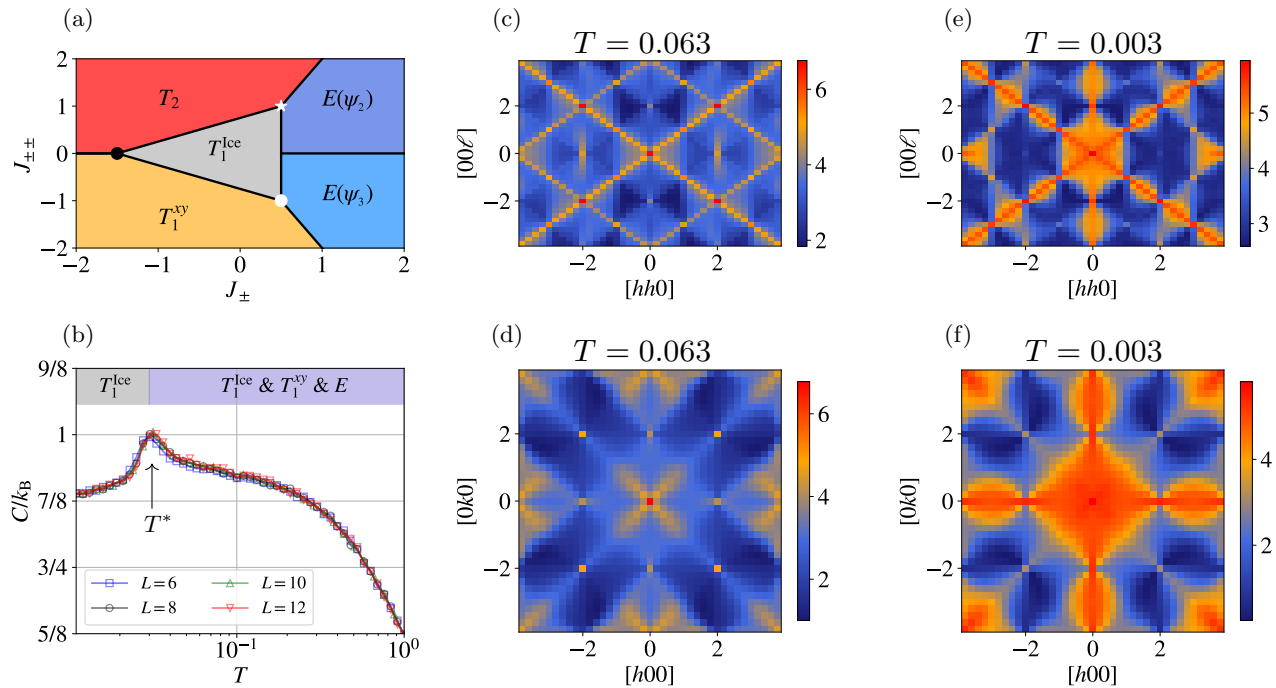
\begin{figure*}[ht!]
\centering
    \begin{tikzpicture}
    \draw (0, 6) node[inner sep=0] {\includegraphics[width=1.\textwidth]{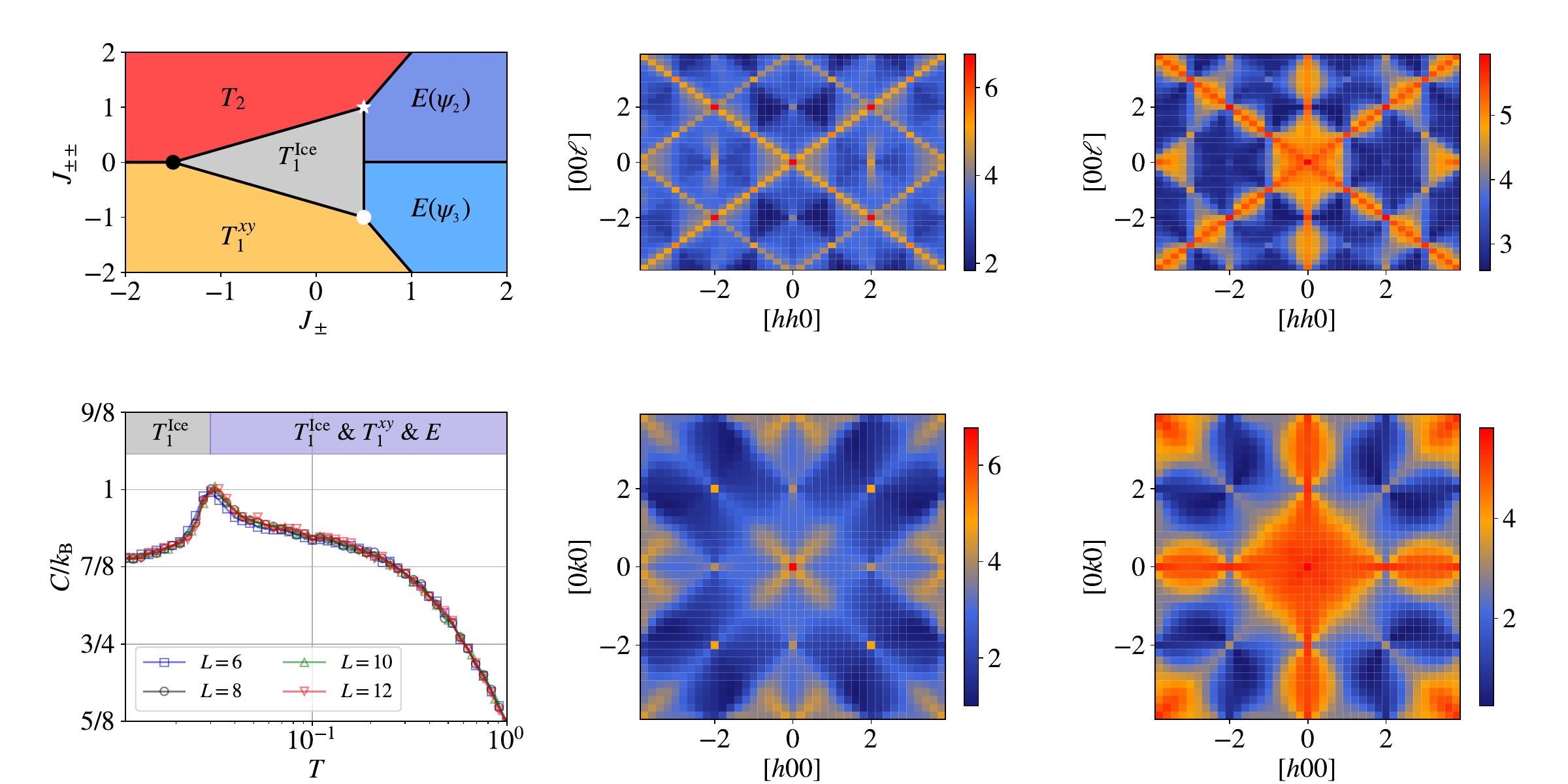}};
    \draw (-7.5, 10.4) node [scale=1.]{(a)};
    \draw (-1.8, 10.4) node [scale=1.]{(c)};
    \draw (4.2, 10.4) node [scale=1.]{(e)};

    \draw (-0.0, 10.1) node [scale=1.3]{$T=0.063$};
    \draw (6, 10.1) node [scale=1.3]{$T=0.003$};

    \draw (-7.5, 6.2) node [scale=1.]{(b)};
    \draw (-1.8, 6.2) node [scale=1.]{(d)};
    \draw (4.2, 6.2) node [scale=1.]{(f)};

    \draw (-0.0, 6.0) node [scale=1.3]{$T=0.063$};
    \draw (6, 6.0) node [scale=1.3]{$T=0.003$};

    \draw (-6.43, 3.7) node [scale=1.]{$T^\ast $};
  
    \coordinate (p1) at (-6.52,3.9);
    \coordinate (p2) at (-6.52,4.5);
    \draw[->] [black] (p1) to (p2);
    \end{tikzpicture}
    \caption{(a) 
    Non-Kramers phase diagram with $J_{zz}=3$ where the DQQ model corresponds to the white dot at the boundary between the  classical $T_1^{xy}$, $E(\psi_3)$, and $T_1^{\rm Ice}$ phases. (b) Specific heat of the DQQ model obtained from Monte Carlo simulations for various system sizes where a bump at a temperature $T^\ast\sim 0.03$ in the specific heat that signals a crossover between an intermediate-temperature and a low-temperature regime, further discussed in the main text, is observed. This crossover is characterized by an entropically driven depopulation of the $E$ and $T_1^{xy}$ irrep modes. Spin structure factors in the $[hh\ell]$ (c)  and $[hk0]$ (d) planes for a temperature just above the crossover temperature $T^\ast$.  Spin structure factors in the $[hh\ell]$ (e) and $[hk0]$ (f) planes for a temperature below the crossover temperature $T^\ast$.}
       \label{fig:fig_1_structure_factors}
\end{figure*}

\section*{Results}

\noindent{\bf Model, irrep analysis and Monte Carlo results.}
The nearest-neighbour spin Hamiltonian, $\mathcal{H}$, for a non-Kramers system is described in terms of three spin-spin coupling constants $\{J_{zz},J_{\pm}, J_{\pm\pm}\}$, 
with $\mathcal{H}$ given by
\begin{eqnarray}
\mathcal{H}&=&\sum_{\langle ij \rangle}J_{zz} S_i^z S_j^z -J_{\pm}( S_i^+ S_j^- + S_i^- S_j^+) \nonumber\\
&& + J_{\pm\pm } (\gamma_{ij}S_i^+ S_j^+ + \gamma_{ij}^\ast S_i^-S_j^-) ,
\label{eq:CH6-H_H+DM} 
\end{eqnarray}
where $S^\alpha_i$ is the $\alpha$th component of the (pseudo) spin-$1/2$ on site $i$ in a local coordinate frame~\cite{SuppMat}, 
with $S_i^z$ representing magnetic dipolar degrees of freedom and $S_i^\pm$ representing electric quadrupolar degrees of freedom~\cite{Lee_kramers_PhysRevB.86.104412,Rau-2019}.
$\langle ij\rangle$ labels the nearest-neighbour pyrochlore bonds between sites $i$ and $j$ and $\gamma_{ij}$ are phase factors imposed by the  lattice symmetry~\cite{Lee_kramers_PhysRevB.86.104412,Rau-2019}.  We take $J_{zz}>0$ to stabilize a spin ice state when $J_{\pm}=J_{\pm\pm}=0$.
To identify the classical ordered phases of ${\cal H}$, we 
first decompose it in terms of the irreducible representations (irreps) of a tetrahedron ~\cite{Yan-2017,our_HDM_paper,SuppMat}. We write
\begin{equation}
    \mathcal{H}=\sum_{\boxtimes} \mathcal{H}^{\boxtimes}
   =\sum_{\boxtimes}\sum_{i,j\in\boxtimes }\bm S_i^T \bm M_{ij} \bm S_j ,
    \label{eq:single-tet-Hamiltonian}
\end{equation}
where $\mathcal{H}^{\boxtimes}$ is the single-tetrahedron Hamiltonian,
\begin{eqnarray}
\mathcal{H}^{\boxtimes}=
\frac{1}{2}
&& \Big [
a_{T_{1}^{\rm Ice}}\left(\bm m^\boxtimes_{T_{1}^{\rm Ice}}\right)^2
+
a_{E}\left(\bm{m}^\boxtimes_{E}\right)^2
+
a_{T_{1}^{xy}}\left(\bm m^\boxtimes_{T_{1}^{xy}}\right)^2
\nonumber\\
&&
+a_{A_2}\left(m^\boxtimes_{A_2}\right)^2
+
a_{T_2}\left(\bm m^\boxtimes_{T_2}\right)^2 
 \Big]\,.
\label{eq:single_tet_Hamiltonian}
\end{eqnarray}
Here, $\{  \bm m^\boxtimes_I  \}$  are the single-tetrahedron irrep spin modes which diagonalize  $\mathcal{H}^{\boxtimes}$. 
The $a_{I}$ parameters are linear functions of the couplings $\{J_{zz},J_{\pm},J_{\pm\pm}\}$ and correspond to the energies associated with irrep $I$~\cite{SuppMat}. In this representation, the $T_1^{xy}$ and the $T_1^{\rm Ice}$ irreps correspond to two splayed ferromagnetic spin configurations, whereas the remaining $A_2,\ E$, and $T_2$ irreps correspond to different antiferromagnetic spin configurations~\cite{Yan-2017}. We refer the reader to the Supporting Information (SI)~\cite{SuppMat} for further details regarding the determination of the $a_I$ parameters and the spin configuration associated with each irrep $I$.

From \eqref{eq:single_tet_Hamiltonian}, the classical ground state phase diagram follows immediately and is shown in Fig.~\ref{fig:fig_1_structure_factors}(a)
~\cite{TakatsuPRL2016TbtiO,Lee_kramers_PhysRevB.86.104412,TaillefumierPhysRevX.7.041057}.
There are four triple points in the phase diagram: 
Three of these correspond to the corners of the gray $T_1^{\rm Ice}$ triangle and one to the intersection of the $T_1^{\rm Ice}$, the  $E(\psi_2)$, and the $E(\psi_3)$ phases. At that point, thermal fluctuations stabilize the $E$ phase which partially invades over the $T_1^{\rm Ice}$, with a transition from paramagnetic to long-range order in the $E$ phase order upon decreasing temperature~\cite{TaillefumierPhysRevX.7.041057}.
We also note that, for the quantum spin-1/2 case, this triple point is near the phase boundary (on the $J_{\pm\pm}=0$ axis where there is a Higgs transition between a quantum spin ice phase and a U(1) magnetic long-range ordered phases~\cite{KatoOnodaPRL2015}.
We do not investigate this $T_1^{\rm Ice}$-$E(\psi_2)$-$E(\psi_3)$ triple point further in the present work.
The leftmost triple point (black circle) corresponds to the local Heisenberg antiferromagnet (HAF) model~\cite{TaillefumierPhysRevX.7.041057}, dual to the highly studied global HAF model~\cite{Rau-2019}, 
latter model known to realize a Coulomb phase spin liquid
~\cite{Moessner98,Moessner-1998b,Henley-2001,ConlonAbsentPhysRevB.81.224413}.

The lower triple point (white circle) is located on the phase boundary between the $E$, the $T_1^{xy}$, the $T_1^{\rm Ice}$ phases. 
This is the DQQ point alluded in the introduction, with the 
fined-tuned parameters $\{J_{zz}=3, J_{\pm}=\frac{1}{2}, J_{\pm\pm}=-1 \}$ defining what we refer to as the DQQ model, and which constitutes the primary focus of the present study. 
The exchange couplings at this specific point can also be parameterized by a global ferromagnetic Heisenberg coupling and the so-called indirect Dzyaloshinskii-Moriya (DM) interaction~\cite{Elhajal05}, namely
 \begin{eqnarray}
\mathcal{H}&=&- J\sum_{\langle ij \rangle}\bm S_i \cdot \bm S_j +D\sum_{\langle ij \rangle}\bm d_{ij}\cdot (\bm S_i \times \bm S_j),
\label{eq:H+DM}
\end{eqnarray}
where $\bm d_{ij}$ are the DM vectors as defined in Ref.~\cite{our_HDM_paper} and the interaction parameters satisfy $D/J=-2$. For more details on the $\bm d_{ij}$ vectors and the sign convention, we refer the reader to the SI. 
Previous work~\cite{our_HDM_paper} noted an apparent lack of magnetic ordering at this triple point.
The last triple point, namely the upper white star in Fig.~\ref{fig:fig_1_structure_factors}(a), 
is at $\{J_{zz}=3, J_{\pm}=\frac{1}{2}, J_{\pm\pm}=1 \}$ --- which we refer to as the DQQ$^*$ model, is dual to the aforementioned DQQ model~\cite{Lee_kramers_PhysRevB.86.104412,Rau-2019}, with both models having identical thermodynamic properties, though their spin-spin correlations differ.

To begin, we first use Monte Carlo (MC) simulations to investigate in detail the thermodynamics and spin-spin correlations of the classical DQQ model, with the spins in \eqref{eq:H+DM} taken as classical vectors of fixed length $\vert {\bm S}_i\vert=1$.
The results are broadly summarized in  Fig.~\ref{fig:fig_1_structure_factors}(b)-(f).
In this figure, and throughout the paper when considering $\mathcal{H}$ in \eqref{eq:H+DM}, as opposed to the more general case of $\mathcal{H}$ in \eqref{eq:CH6-H_H+DM}, we fix $D=-2J$ and measure all energies in units of $J$ and temperature in units of $J/k_{\textrm B}$. 
Figure~\ref{fig:fig_1_structure_factors}(b) shows the temperature dependence of the specific heat, $C$, of the DQQ model, which
 exhibits a fairly sharp peak at temperature $T^\ast\sim O(J \times 10^{-2})$ [note the logarithmic temperature scale in 
 Fig.~\ref{fig:fig_1_structure_factors}(b)],  and asymptotically plateaus to a value of $C/k_{\rm B}\approx 7/8$ at the lowest temperature considered.
Figure~\ref{fig:fig_1_structure_factors}(c)-(f) illustrates the evolution of the spin structure factors \emph{transverse} to wave vector $\bm q$, as probed by unpolarized neutron scattering~\cite{SuppMat,Fennelscience.1177582, Morris_science,ChungKristianPhysRevLett.128.107201}, in the $[hh\ell]$ and $[hk0]$ planes for $T>T^\ast$ in subpanels (c) and (d), and for $T<T^\ast$ in subpanels (e) and (f). 
Here, we consider an isotropic (diagonal and unitary) $g$-tensor to obtain what we refer to as the ``spin structure factor''~\cite{SuppMat}.
This is in line with the approach taken in Ref.~\cite{Kadowaki_2015} to expose the various features displayed by correlation functions which originate from the intertwinned magnetic dipolar and electric quadrupolar degrees of freedom represented by
the pseudospin components $S^{z}$ and $S^\pm$,
respectively~\cite{Lee_kramers_PhysRevB.86.104412}.
We shall discuss the experimental implications of considering the true $g$-tensor on the neutron structure factor of non-Kramers ion systems below in the Discussion section.

The spin-spin correlations of the DQQ model display a plethora of rich anisotropic features in ${\bm q}$-space, which are further discussed below. However, perhaps most interesting is their very 
rapid change when the system passes from above to below $T^\ast$. As we show next, the anisotropic features for $\cramped{T > T^\ast}$ can be understood by considering a long-wavelength theory composed of competing rank-1 and rank-2 tensor fields, both of whose low-temperature behaviour is constrained by an emergent Gauss' law. The change in correlation functions at $T^\ast$ is associated with a spin liquid to spin liquid crossover driven by entropic effects. Additionally, we will show later that this novel intertwined rank-1 ($R_1$) and rank-2 ($R_2$), $R_1$-$R_2$, spin liquid phase appears to be stable even in the quantum spin-$1/2$ case, as suggested by pseudo-fermion functional renormalization group (PFFRG) calculations. 
\\

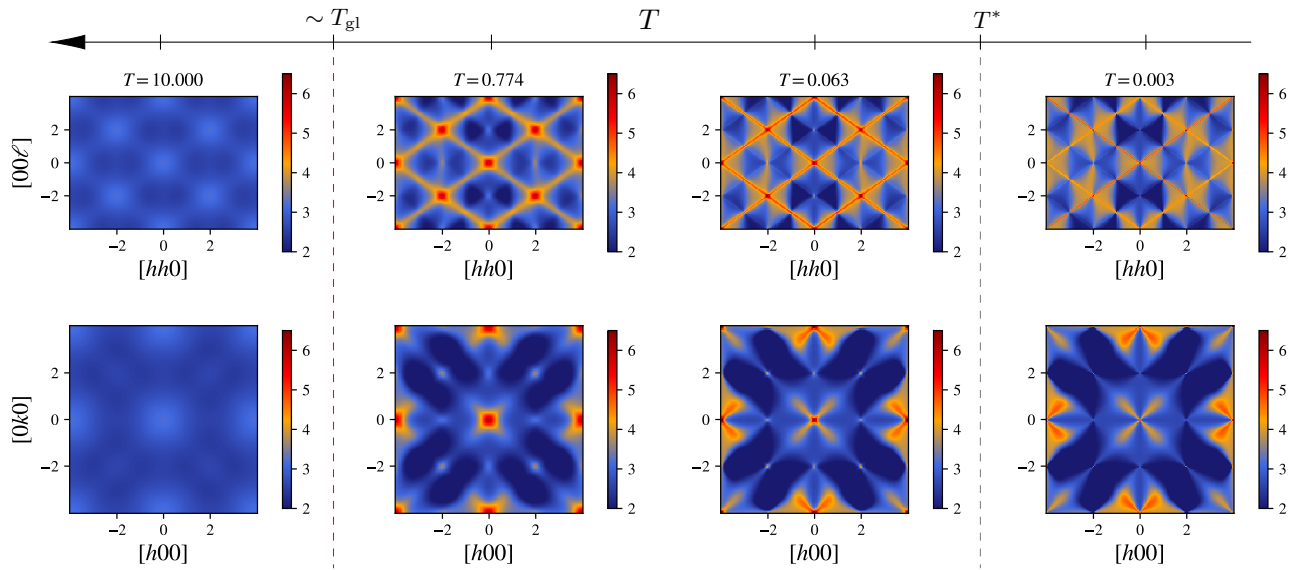
\begin{figure*}[ht!]
\centering

    \begin{tikzpicture}
    \draw (0, 6) node[inner sep=0] {\includegraphics[width=1.\textwidth]{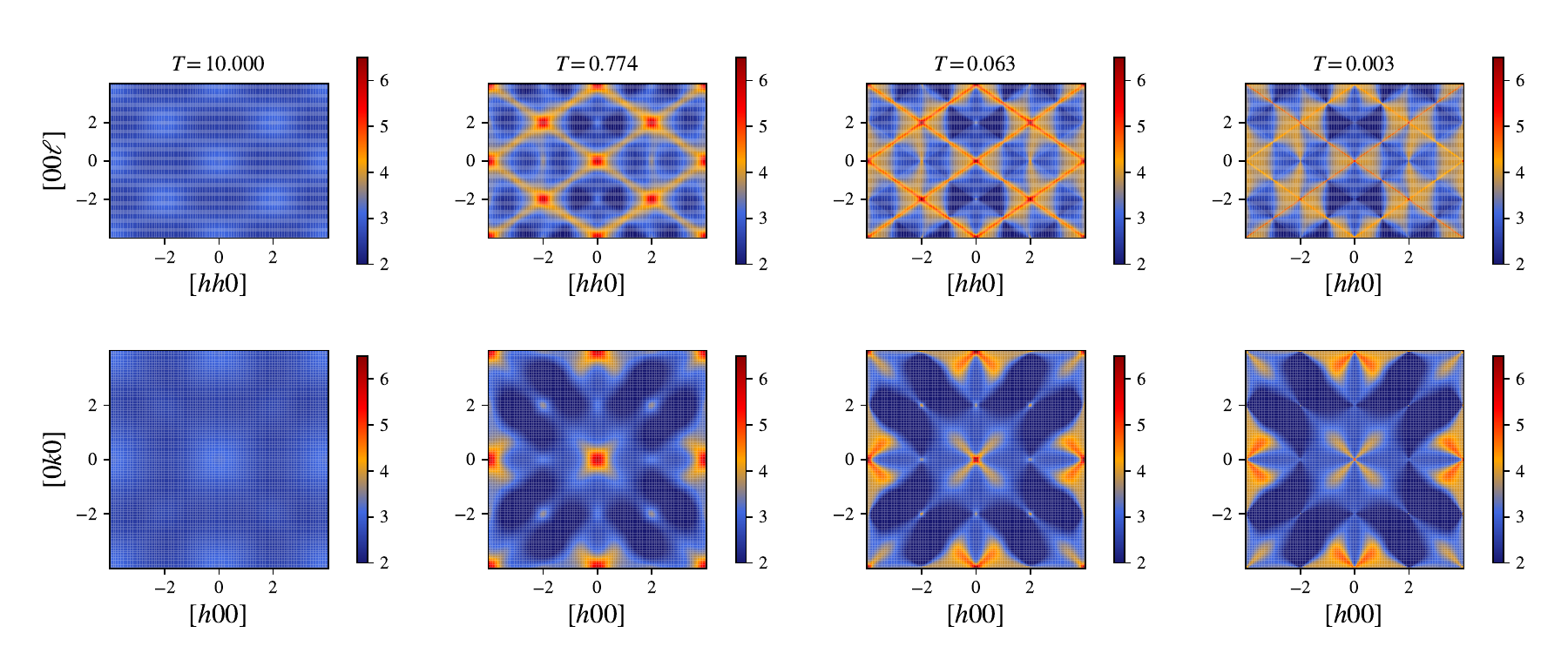}};


    \draw (4.5, 9.8) node [scale=1.]{$T^\ast $};

     \draw (-4.2, 9.8) node [scale=1.]{$\sim T_{\rm gl} $};
     \coordinate (p1) at (-4.2,9.6);
    \coordinate (p2) at (-4.2,9.4);
    \draw[-] [black] (p1) to (p2);
     \coordinate (p1) at (-4.2,9.3);
    \coordinate (p2) at (-4.2,2.5);
    \draw[dashed] [red] (p1) to (p2);
    
    \draw (0, 9.8) node [scale=1.2]{$T $};

    
    \coordinate (p1) at (-6.5,9.6);
    \coordinate (p2) at (-6.5,9.4);
    \draw[-] [black] (p1) to (p2);

    \coordinate (p1) at (-2.1,9.6);
    \coordinate (p2) at (-2.1,9.4);
    \draw[-] [black] (p1) to (p2);

    \coordinate (p1) at (2.2,9.6);
    \coordinate (p2) at (2.2,9.4);
    \draw[-] [black] (p1) to (p2);

    \coordinate (p1) at (6.6,9.6);
    \coordinate (p2) at (6.6,9.4);
    \draw[-] [black] (p1) to (p2);

    \coordinate (p1) at (4.4,9.6);
    \coordinate (p2) at (4.4,9.4);
    \draw[-] [black] (p1) to (p2);
  
    \coordinate (p1) at (4.4,9.3);
    \coordinate (p2) at (4.4,2.5);
    \draw[dashed] [gray] (p1) to (p2);

    \draw[{Latex[length=5mm, width=2mm]}-] (-8,9.5)--(8,9.5) node[right]{};
    \end{tikzpicture}
    \caption{Spin structure factors in the $[hh\ell]$ plane (upper row) and $[hk0]$ plane (lower row), obtained using a self-consistent Gaussian approximation for the temperatures illustrated above each column of figures. The red dashed line separates the gas-like (paramagnetic) regime at $T\gtrsim T_{\rm gl}$ from the intermediate liquid regime at $T^\ast<T\lesssim T_{\rm gl}$. The gray line at $T^\ast$ corresponds to the peak temperature of the heat capacity shown in Fig.~\ref{fig:fig_1_structure_factors}(a).}
    \label{fig:fig_1_SCGA}
\end{figure*}

\noindent{\bf Self-consistent Gaussian approximation (SCGA) and effective long-wavelength theory.}
The low energy configurations of the DQQ model are built from three out of the five irrep modes~\cite{our_HDM_paper}, see SI for more details. 
To further elaborate on the degeneracy of the ground-state manifold,  
we Fourier transform the interaction matrix in Eq.~(\ref{eq:single-tet-Hamiltonian}) to obtain $\bm M(\bm q)$.
The spectrum of this matrix displays four degenerate low-energy flat bands~\cite{SuppMat} which suggests an extensively degenerate ground state manifold \textendash~a telltale sign of a CSL~\cite{Benton-2021,yan2023classification}. 
To construct an effective low-energy theory,
we apply a self-consistent Gaussian approximation (SCGA), 
an approach proven useful in previous studies of 
CSLs~\cite{Garanin,Isakov04,ConlonAbsentPhysRevB.81.224413,ChungKristianPhysRevLett.128.107201}.

The spin correlation functions obtained through an SCGA analysis are shown in Fig.~\ref{fig:fig_1_SCGA} for various temperatures, and for both the $[hh\ell]$ and the $[hk0]$ scattering planes. These SCGA structure factors 
show a progressive evolution from high-temperatures ($T\gg T_{\rm gl}$, with $T_{\rm gl}$ the paramagnetic (``spin gas'') to spin liquid crossover temperature), where the correlation function is nearly featureless in the paramagnetic regime, to the lower temperature spin liquid regime (e.g. $T\sim 0.774$), where sharp features have become visible.
A direct comparison between Fig.~\ref{fig:fig_1_SCGA} and Fig.~\ref{fig:fig_1_structure_factors}(c) and (d) shows that the SCGA captures the anisotropic features observed in our MC simulations below $T_{\rm gl}$ and \textit{above} $T^\ast$, but not below $T^\ast$. Twofold~\cite{Isakov04} and fourfold pinch points~\cite{YanRank2U1PhysRevLett.124.127203,Benton-2021} as well as continuous lines of scattering, dubbed pinch-lines~\cite{Benton2016Pinch-line-singularity,Niggemann-2023}, can be seen in both the MC and SCGA results. More specifically, in the $[hh\ell]$ plane in Fig.~\ref{fig:fig_1_structure_factors}(c) twofold pinch points (at $[hh\ell]=[220],[222],[002]$ and symmetry related points) as well as lines of strong scattering intensity along the $[111]$ and $[\bar{1}\bar{1}1]$ directions are observed. 
Additionally, fourfold pinch points at $[hk0]=[000]$ are seen in the $[hk0]$ plane in Fig.~\ref{fig:fig_1_structure_factors}(d). 
For more details, see the SI~\cite{SuppMat}. 

The observation of such anisotropic ${\bm q}$-space features had previously been related to underlying emergent gauge symmetries. Twofold pinch points are indicative of a  divergence-free constraint for a vector field~\cite{Henley-2010}, while the fourfold pinch points and lines  are related to the emergence of a rank-2 tensor field 
with an associated Gauss' law constraint~\cite{YanRank2U1PhysRevLett.124.127203,Benton2016Pinch-line-singularity}. The agreement between the SCGA and the MC structure factors \textit{above} $T^\ast$ motivates the construction of an effective long-wavelength theory that we present next. On the other hand, the clear discrepancy between the SCGA and the MC structure factors for $T<T^\ast$ implies that the behaviour in the lowest temperature regime requires a theory that goes beyond SCGA, which we shall discuss later.

\begin{figure*}[t]
\centering
    \begin{overpic}[width=\textwidth]{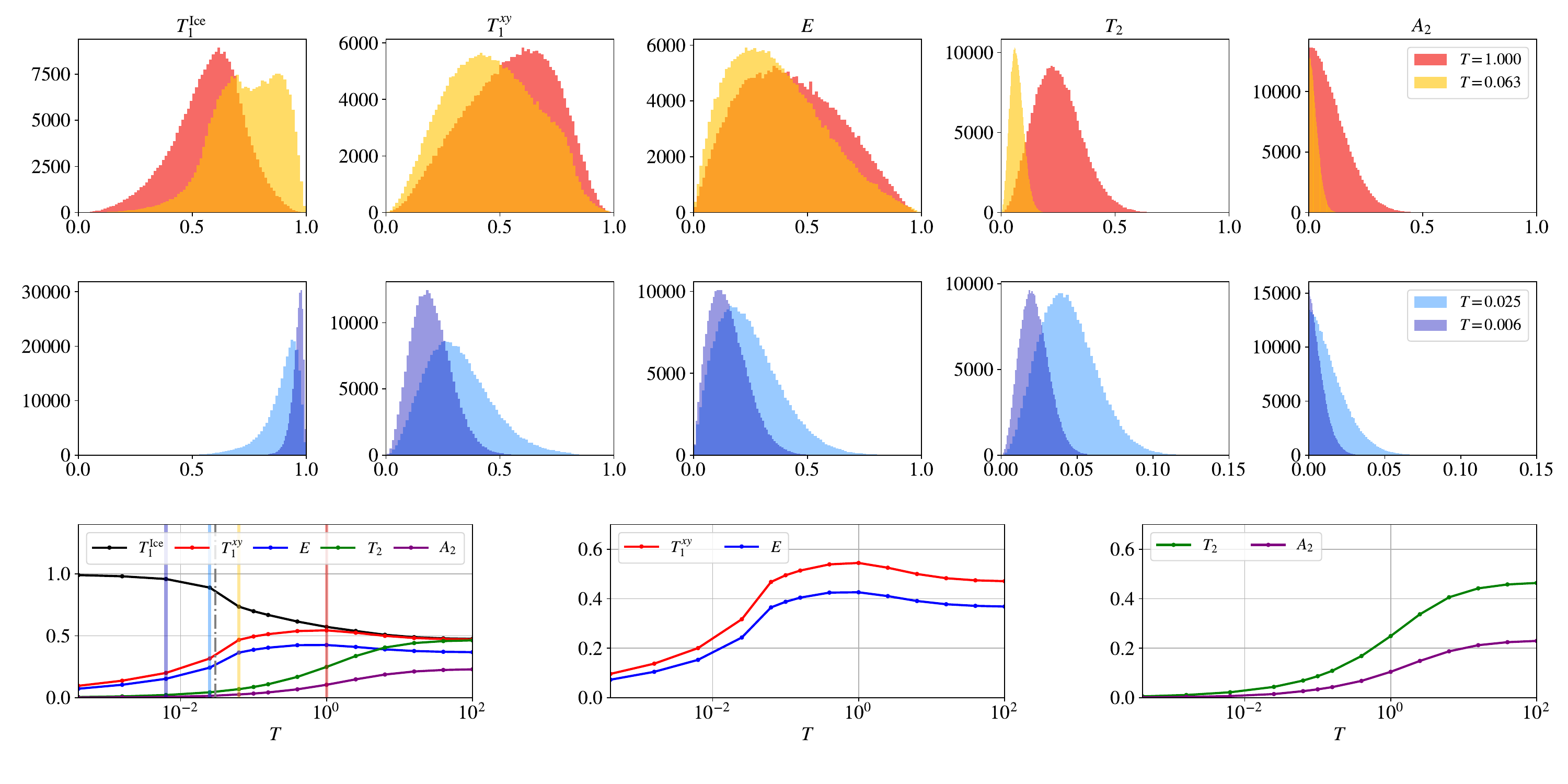}
    \put(3,49) {(a)}
    \put(22,49) {(b)}
    \put(42,49) {(c)}
    \put(62,49) {(d)}
    \put(82,49) {(e)}

    \put(3,33) {(f)}
    \put(22,33) {(g)}
    \put(42,33) {(h)}
    \put(62,33) {(i)}
    \put(82,33) {(j)}

    \put(3,17.5) {(k)}
    \put(37,17.5) {(l)}
    \put(71,17.5) {(m)}

    \put(10.15,17) {\textcolor{violet}{$\downarrow$}}
    \put(12.9,17) {\textcolor{blue}{$\downarrow$}}
    
    \put(12.8,4.6)  {\fontsize{6}{38} $\uparrow$}
    \put(12.8,3.3)  {\fontsize{6}{38} $T^\ast$}
    
    \put(20.5,11.3)  {\fontsize{6}{38} $\leftarrow$}
    \put(22,11.3)  {\fontsize{6}{38} $\sim T_{\rm gl}$}

    \put(14.83,17) {\textcolor{orange}{$\downarrow$}}
    \put(20.4,17) {\textcolor{red}{$\downarrow$}}
    \end{overpic}
    \caption{ (a)-(j) Distribution of the irrep mode magnitude of all the \textit{up} tetrahedra $\{|\bm m_{I}^{\widehat{\boxtimes}}|\}$ of 250 Monte Carlo sampled configurations for a system size $L=10$ at the temperatures indicated in the rightmost panel. Each column of panels (a)-(j) corresponds to the distribution of a given irrep, namely, from left to right, the $T_1^{\rm Ice}$, the $T_1^{xy}$, the $E$, the $T_2$, and the $A_2$ irreps. (k)-(m) Evolution of the average irrep mode magnitude, $\frac{1}{L^3}\langle \sum_{\widehat{\boxtimes}} |m_I^{\widehat{\boxtimes}}|\rangle$, on a lattice as a function of temperature averaged over $2500$ configurations sampled through Monte Carlo. The vertical shaded lines indicate the temperatures where the configurations used to produce the histograms in the first and second rows were sampled from. In subpanel (k), we identify the gray 
    dashed-dotted line with the liquid-to-liquid crossover temperature $T^\ast$, and the red vertical line at the temperature $T=1.0$, with the approximate liquid-to-gas crossover temperature $T_{\rm gl}$. We note that at low temperatures the average irrep mode magnitude corresponding to the $T_1^{\rm Ice}$ irrep approaches unity elucidating a selection of configurations with this irrep mode over configurations composed of the other two low-energy degenerate $xy$ 
    ($T_1^{xy}$ and $E$) irreps.}
    \label{fig:fig_2_statistics_MC}
\end{figure*}

To begin, we note that Monte Carlo simulations reveal that the two  high-energy  $A_2$ and $T_2$ irreps begin to depopulate at $T\sim O(J)$ [see Fig.~\ref{fig:fig_2_statistics_MC}(d)-(e)], leaving only the degenerate low-energy  $T_1^{\rm Ice}$,  $T_1^{xy}$ and $E$ irreps [see Fig.~\ref{fig:fig_2_statistics_MC}(a)-(c)] 
thermally populated and relevant for $T\lesssim O(J)$.
Therefore, and similarly to the approach taken in Refs.~\cite{Benton2016Pinch-line-singularity,YanRank2U1PhysRevLett.124.127203}, we proceed to construct an effective theory for $T>T^\ast$  that solely focuses on the  $T_1^{\rm Ice}$,  $T_1^{xy}$ and $E$ irreps in the  temperature range $T^* \lesssim T \lesssim  T_{\textrm {gl}}$. 
Therefore, starting from the Hamiltonian in \eqref{eq:CH6-H_H+DM}, we define a rank-1 field, 
\begin{eqnarray}
(\bm B^{\rm Ice})^\alpha \equiv m_{T_1^{\rm Ice}}^\alpha
\label{eq:fields},
\end{eqnarray}
and a rank-2 field 
\begin{align}\label{eq:fields_2}
&\small
\mathcal{M}^{xy} 
\!\!\equiv\\& \!\! 
\sqrt{\frac{2}{3}}
\begin{pmatrix}
\cramped{\frac{\sqrt{3}}{2}m_{\psi_2}-\frac{1}{2}m_{\psi_3}}& -\frac{\sqrt{3}}{2} m_{T_1^{xy}}^{z}&  \frac{\sqrt{3}}{2} m_{T_1^{xy}}^{y} \\
\frac{\sqrt{3}}{2} m_{T_1^{xy}}^{z} & \cramped{-\frac{\sqrt{3}}{2}m_{\psi_2}-\frac{1}{2}m_{\psi_3}} & -\frac{\sqrt{3}}{2} m_{T_1^{xy}}^{x}\\
-\frac{\sqrt{3}}{2} m_{T_1^{xy}}^{y} &\frac{\sqrt{3}}{2} m_{T_1^{xy}}^{x}& m_{\psi_3}
\end{pmatrix} .
\nonumber 
\normalsize
\end{align}
Here, $\bm B^{\rm Ice}$ corresponds to the fluxes in the Coulomb phase~\cite{Castelnovo-2012} (components of $T_1^{\rm Ice}$), while $m_{T_1^{xy}}^{\alpha}$ are components of $T_1^{xy}$, while $ m_{\psi_2}$ and $ m_{\psi_3}$ are components of the $E$ irrep~\cite{SuppMat}. 
In passing, we note that the rank-1 field $\bm B^{\rm Ice}$ and rank-2 $\mathcal{M}^{xy}$ field are uniquely composed by the local-$z$ and local-$xy$ degrees of freedom of the spins, respectively.
With these, the long-wavelength approximation to the SCGA Hamiltonian  reads~\cite{SuppMat}  
\begin{eqnarray}
\beta\mathcal{H}&=&\beta E_0+ \frac{3}{16}\beta J\int d^3 \bm q \left(|\bm{q}\cdot \bm{B}^{\rm Ice}|^2 + 
|\bm{q}^T \mathcal{M}^{xy}|^2\right) 
\label{eq:eff_theory}
\\&&
+ \lambda\int d^3 \bm q  
\left( 
| \bm{B}^{\rm{Ice}}|^2 + 
\mathrm{Tr}\left[(\mathcal{M}^{xy})^T \mathcal{M}^{xy}\right] 
\right) 
+ O(\bm{q}^4).
\nonumber 
\end{eqnarray}
This Hamiltonian consists of three terms: the first term, $E_0$, denotes the ground state energy of the three degenerate irrep modes. The second term is composed of \textit{two} emergent Gauss' laws: one for the vector field $\bm{B}^{\rm Ice}$ and another for the rank-2 field $ \mathcal{M}^{xy}$, which in the limit $T\rightarrow0$ correspond to $\bm q \cdot  \bm{B}^{\rm Ice}=0$ and $\bm q^T   \mathcal{M}^{xy}=0$  and, in direct space, $\nabla\cdot \bm{B}^{\rm Ice} \equiv 
\partial_\alpha  (\bm{B}^{\rm Ice})^\alpha=0$ and 
$\partial_\alpha (\mathcal{M}^{xy})^{\alpha,\beta}=0$, with an implicit sum over repeated indices. The third term proportional to $\lambda$ corresponds to the spin-length constraint in the SCGA approximation~\cite{SuppMat}. The terms corresponding to the local-$z$ spin components, i.e., those with the vector field $\bm{B}^{\rm Ice}$, describe an effective Coulomb phase where a divergence-free condition results in the twofold pinch points~\cite{Henley-1989,Moessner-1998b}. The terms describing the  divergence-free condition of  the local-$xy$ components, i.e. on the   $\mathcal{M}^{xy}$ tensor field, lead to fourfold pinch points in the $[hk0]$ plane~\cite{PreM_PhysRevB.98.165140}. 
Thus, the full set of anisotropic features observed in the spin structure factors along with the effective theory in \eqref{eq:eff_theory} imply the existence of two sets of emergent gauge fields constrained by respective Gauss' law.
We thus identify the spin liquid state in the intermediate temperature regime 
$\cramped{ T^* \lesssim T \lesssim  T_{\textrm {gl}} }$ as a novel rank-1 -- rank-2 (henceforth $R_1$-$R_2$) spin liquid.

For $T<T^\ast$, some of the anisotropic features in the spin structure factor disappear entirely, implying that a different long-wavelength theory is needed there. We note that spin structure factors for $T<T^\ast$ are spin-ice-like where only twofold~\cite{Isakov04,ChungKristianPhysRevLett.128.107201} pinch points 
are seen. Hence, we expect the $T<T^\ast$ theory to resemble the Coulomb phase theory with a single Gauss' law on a (rank-1) vector field~\cite{Castelnovo-2012,ChungKristianPhysRevLett.128.107201}.
Indeed, such a theory will result from \eqref{eq:eff_theory} \emph{if} the $\mathcal{M}^{xy}$ tensor field, and consequently the $T_1^{xy}$ and $E$ irreps, were to thermally depopulate (freeze-out) due to missing higher-order gradient terms in the effective Hamiltonian which become non-negligible at lower temperatures. 
As way of confirming this expectation, the depopulation of 
the $T_1^{xy}$ and $E$ irreps for $T<T^\ast$ can be seen in the MC data of Fig.~\ref{fig:fig_2_statistics_MC}(g)-(h), 
with the distribution of all irrep projections below $T^\ast$ shown in Fig~\ref{fig:fig_2_statistics_MC}(f)-(h).
The depopulation of $T^{xy}_{1}$ and $E$ irreps implies that low-temperature spin configurations for $T<T^\ast$ are solely made up of $T_1^{\rm Ice}$, as observed in Fig.~\ref{fig:fig_2_statistics_MC}(f).
As a summary, we present in Fig~\ref{fig:fig_2_statistics_MC}(k)-(m) the average value  of each irrep projection 
as a function of temperature. To summarize, two regimes below $T_{\textrm{gl}}$, separated by the crossover temperature $T^\ast$, can be identified: one where $T^{\rm Ice}_{1}$,  $T^{xy}_{1}$, and $E$ irreps have similar values ($T^\ast \lesssim T \lesssim T_{\rm gl}$), and another ($T \lesssim T^\ast$) where all irreps but $T^{\rm Ice}_{1}$ are thermally depopulated.

We now discuss the mechanism for the depopulation of $T^{xy}_{1}$, and $E$ irreps, and the thermal crossover between the two spin liquid phases. 
The value of the specific heat plateau at low temperatures in Fig.~\ref{fig:fig_1_structure_factors}(b), namely, $C/k_{\rm B}\sim 7/8=0.875$, and not $1$ as expected from the equipartition theorem, already hints at what is the mechanism at play~\cite{Moessner-1998b,ChalkerHoldsworthKagomePhysRevLett.68.855}. Specifically, 
it suggests that low-energy excitations above the ground state are not all quadratic, but that higher-order fluctuation  modes are being excited. 
To confirm this, we investigate the low energy spin fluctuations about individual ground state configurations through a classical low-temperature expansion (CLTE). 
\\

\noindent{\bf Classical low-temperature expansion: entropic selection of the rank-1 spin liquid.}
By borrowing the method from Ref.~\cite{Walker_Walstedt_1980}, we numerically construct a CLTE which yields a quadratic theory of spin fluctuations in \textit{real-space} about a ground-state configuration. 
The spin-fluctuation modes and their corresponding energies are identified as the eigenmodes and eigenvalues of the (quadratic spin-fluctuation) Hessian matrix~\cite{SuppMat}.
In this approach, quartic and higher-order modes show up as zero-modes as the quadratic CLTE theory does not contain higher-order spin fluctuations terms~\cite{SuppMat}. This observation therefore allows us to quantify the fraction of higher-order spin-fluctuating modes by tracking the number of quadratic zero-modes identified in this theory.

As discussed above, the depopulation of the $E$ and the $T_1^{xy}$ irreps 
and the increasing population  of the $T^{\rm Ice}_{1}$ irrep below $T^\ast$ 
implies the selection of spin-ice configurations at low temperatures, see Fig.~\ref{fig:fig_2_statistics_MC}.
To expose the driving mechanism behind this selection, we apply the real-space CLTE to two types of spin configurations:
(i) to various \textit{pure} spin-ice configurations 
(where all spins are constrained to point along their local $z$ axis), and (ii) to numerically obtained 
non-spin-ice ground state configurations (where the spins are not constrained to point along the local-$z$ axis). These two sets of configurations are obtained by employing classical Monte Carlo simulations on an Ising antiferromagnetic model for the spin-ice configurations, and a numerical iterative energy minimization (IM) directly at $T=0$ 
(obtained with the maximal numerical accuracy that we could afford) for the non-spin-ice configurations~\cite{SuppMat}. Both the IM and the spin-ice configurations are obtained for a system of linear size $L=10$.

\begin{figure}[ht!]
\centering
    \begin{overpic}[width=.8\linewidth]{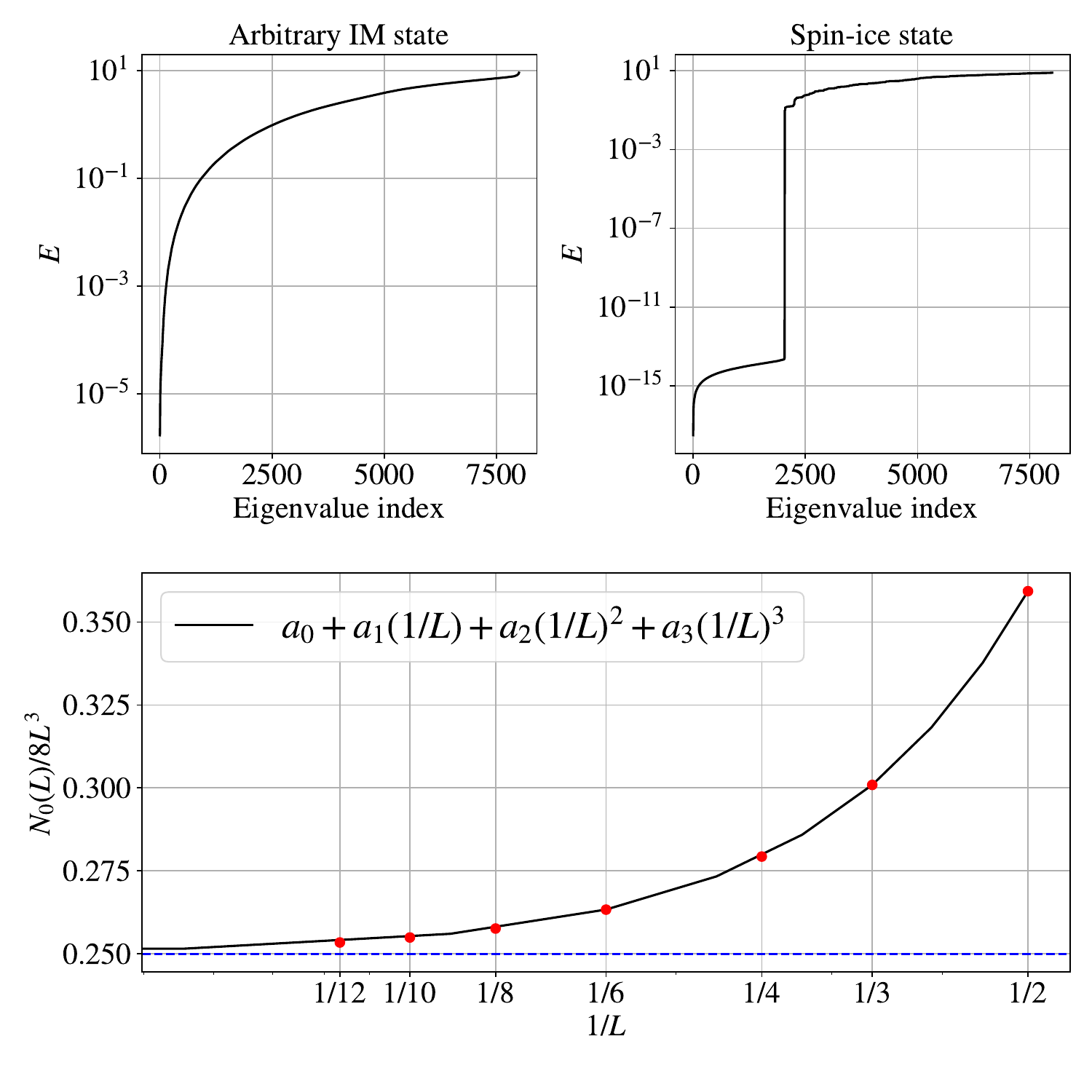}
    \put(10,98){(a)}
    \put(58,98){(b)}
    \put(10,50){(c)}
    \end{overpic}
    \caption{ Energy eigenvalues 
    of the spin fluctuation modes obtained from a real space CLTE where the ground-state configurations are a minimum energy configuration obtained through IM (a), and a spin-ice configuration obtained via classical Monte Carlo on an Ising AFM model (b), both states where obtained for a system size of $L=10$. In these plots, the $x$ axis labels the numbered eigenvalue index, ranging from $1$ to $8000$, and the $y$ axis labels the energy of the eigenvalues. (c) Evolution of the fraction of zero-mode eigenvalues about a spin-ice configuration for system size $L\in \{2,3,4,6,8,10,12\}$ averaged over $100$ spin-ice configurations. Here the black line is a third order in $1/L$ polynomial fit to the data where $a_0\sim 0.25$. }
    \label{fig:fig_3_CLTE}
\end{figure}

Figures~\ref{fig:fig_3_CLTE}(a) and (b) depict the energy eigenvalues obtained from a CLTE where the starting ground-state is (a) a configuration obtained via IM and (b) perfect spin-ice configuration. 
For the configuration obtained via IM, the
energy spectrum in Fig.~\ref{fig:fig_3_CLTE}(a) shows a continuous progression of energy eigenvalues with the smallest eigenvalues of order $O(J\times 10^{-5})$. In contrast, for the spin-ice configuration, the energy spectrum illustrated in Fig.~\ref{fig:fig_3_CLTE}(b) shows a sudden drop and a significant fraction of eigenvalues with energy below $O(J\times 10^{-11})$.
We identify these as zero-modes within the quadratic theory. The comparison of the two spectra along with the monotonous thermal depopulation of the $E$ and $T_1^{xy}$ irreps indicate that spin fluctuations about a spin-ice configuration are softer than those about a non-spin-ice configuration
implying that spin-ice configurations possess a lower free-energy than the non-spin-ice states obtained through IM, resulting in their selection at low temperatures. 
Although the spectra presented in Figs.~\ref{fig:fig_3_CLTE}(a) and (b) are only shown for two unique configurations, we have carried out the analysis for about $100$ different IM and pure spin-ice configurations and obtained quantitatively equivalent energy spectra distributions.

To find the fraction of quartic modes about a spin-ice configuration, we count the number of vanishing energy eigenvalues (we identify an eigenvalue with a zero-mode if its numerical value is below $J\times 10^{-9}$) and study how the fraction of these evolves as a function of system size $L$. 
We fit the number of zero eigenvalues as a function of $L$ using the following form
\begin{eqnarray}
    N_0(L)=8L^3(a_0+a_1/L+a_2/L^2+a_3/L^3),
    \label{eq:fraction_modes}
\end{eqnarray}
where $a_0,\ a_1,\ a_2,\ a_3$ correspond to the fraction of zero eigenvalues originating from local, one-dimensional, two-dimensional, and global zero modes, respectively. 
We consider system sizes $L=\{2,3,4,6,8,10,12\}$, calculate the fraction of zero eigenvalues, i.e., $N_0(L)/8L^3$, and plot it as a function of $L$. 
As seen in Fig~\ref{fig:fig_3_CLTE}(c), the fraction of zero eigenvalues approaches very precisely $1/4$ as the system size approaches the thermodynamic limit, $1/L\to 0$.

This fraction of zero quadratic modes is independently confirmed by the aforementioned observed value of the specific heat $C/k_{\rm B}\approx 7/8$ at low temperatures. This result suggests that $3/4$ of the modes are quadratic whereas the remaining $1/4$ are quartic, with $C=(n_2k_{\rm B}/2+n_4k_{\rm B}/4)/N_{\rm s}$, where $N_{\rm s}=4$ is the number of spins per tetrahedron, and $n_2=6$ and $n_4=2$ are numbers of quadratic and quartic modes per tetrahedron, respectively~\cite{Moessner-1998b,ChalkerHoldsworthKagomePhysRevLett.68.855}. 

To summarize, our MC analysis shows no evidence for a symmetry-breaking transition down to zero temperature. Rather, it identifies a novel $R_1$-$R_2$ spin liquid state that develops upon cooling from the paramagnetic phase and which is followed upon further cooling by a crossover to a spin-ice-like spin liquid at a temperature $T^\ast$.
The CLTE analysis implies that the thermal crossover at the temperature $T^\ast$ and, therefore, the selection of the low-temperature spin-ice-like spin liquid in the DQQ model proceeds via an entropic mechanism. While this mechanism is similar to an order-by-disorder selection of a magnetically ordered state in a degenerate manifold~\cite{our_HDM_paper,Javanparast15}, for the DQQ model, the selection does not result in a magnetically ordered state. 
To the best of our knowledge, this finding constitutes the first observation of a classical temperature-driven ``disorder-by-disorder'' mechanism where an extensively degenerate sub-manifold is selected by thermal fluctuations.
\\

\begin{figure*}[t!]
    \centering
       \begin{subfigure}[b]{0.52\textwidth}
        \centering
        \begin{overpic}[width = 1\textwidth]{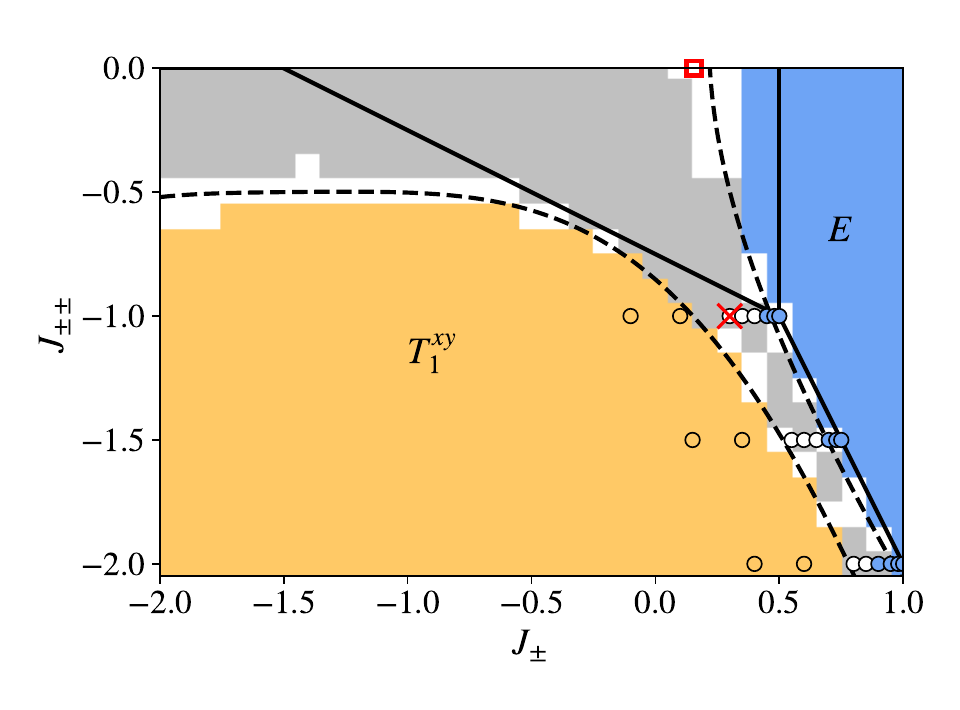}
        \put(0,68){(a)}
        \end{overpic}
        \vspace{-5mm}
    \end{subfigure}
    \begin{subfigure}[b]{0.45\textwidth}
        \centering
        \begin{overpic}[width = 1\textwidth]{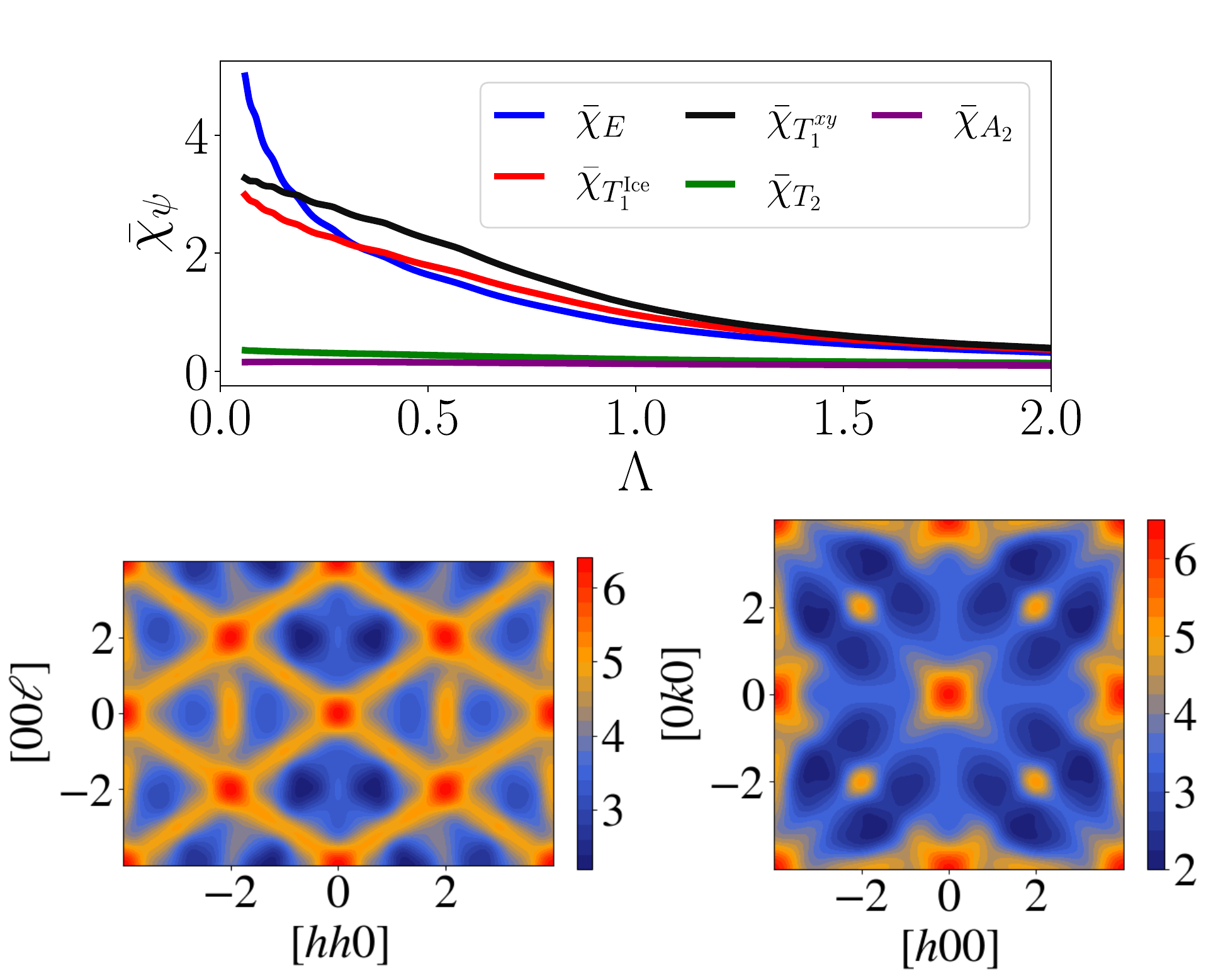}
            \put(2,77){(b)}
            \put(2,35){(c)}
            \put(54,35){(d)}
        \end{overpic}
    \end{subfigure}
    \caption{ (a) PFFRG phase diagram of the spin-$1/2$ non-Kramers pyrochlore model at $T=0$ with fixed $J_{zz}=3$ where the gray region denotes an absence of magnetic long-range order, the yellow and blue regions correspond to the $\bm q=\bm 0$ quadrupolar orders $T_1^{xy}$ and $E$, respectively, and the white regions are of uncertain magnetic behaviour. As a guide to the eye, the approximate quantum phase boundaries from PFFRG are indicated by dashed lines. Solid black lines mark the classical phase boundaries which meet at the DQQ point  
    ($J_{zz}=3$, $J_{\pm}=\frac{1}{2}$ and $J_{\pm\pm}=-1$). The HTSE results are shown as circles whose colors correspond to the order parameter susceptibility dominating in a calculation up to order 
    $1/T^8$. The phase boundary between quantum spin ice and magnetic $E$ phases, as previously determined by quantum Monte Carlo on the unfrustrated $J_{\pm\pm}=0$-line (the so-called XXZ model) is marked by a red square~\cite{KatoOnodaPRL2015}.
    (b) Order parameter susceptibilities $\bar{\chi}_{\psi}$ from PFFRG at $T=0$ as a function of the renormalization group parameter $\Lambda$ for the quantum spin-$1/2$ model with interactions $\{J_{zz},J_{\pm},J_{\pm\pm}\} = \{3.0,0.3,-1.0\}$, marked by a red cross in the phase diagram in (a).
       (c), (d) Static (zero frequency) spin structure factors from PFFRG at $T=0$ for the same model as in (b) within the $[hh\ell]$ and $[hk0]$ planes in the low-cutoff limit $\Lambda\rightarrow0$.}
    \label{fig:Q_HTSE+PFFRG}
\end{figure*}

\noindent{\bf Quantum spin-$1/2$ model. } 
We now turn to the corresponding quantum spin-$1/2$ version of the DQQ model to investigate how quantum fluctuations modify our findings for the classical system. We perform this study in an extended parameter regime of the exchange constants $J_{zz}$, $J_{\pm}$, $J_{\pm\pm}$, varying them in the vicinity of  the DQQ point 
$\cramped{(J_{zz}=3, J_{\pm}=\frac{1}{2},J_{\pm\pm}=-1)}$.
This non-Kramers Hamiltonian with positive $J_{zz}$ has been studied before by both quantum and classical methods~\cite{KatoOnodaPRL2015,Lee_kramers_PhysRevB.86.104412,Onoda_Kramers_PhysRevB.83.094411,TaillefumierPhysRevX.7.041057,TakatsuPRL2016TbtiO}. The classical model at $T=0$ shows the four phases of  Fig.~\ref{fig:fig_1_structure_factors}(a) (three quadrupolar phases associated with the irreps $E$, $T_2$, $T_1^{xy}$ and one dipolar phase with the irrep $T_1^{\rm Ice}$). The most striking observation in studies of the quantum spin-$1/2$ model is the replacement of the classical spin-ice phase by a U(1) quantum spin liquid in the vicinity of the Ising point $J_{\pm}=J_{\pm\pm}=0$~\cite{Gingras_2014,KatoOnodaPRL2015}.

We first apply the PFFRG method~\cite{Reuther10,Mueller23} to compute the $T=0$, spin-$1/2$ phase diagram of the non-Kramers model, shown in Fig.~\ref{fig:Q_HTSE+PFFRG}(a), focusing on the parameter region around the DQQ point. This method determines magnetic ordering via the presence of an RG flow breakdown of the cutoff-dependent susceptibility~\cite{SuppMat}. The nature of the magnetic order is further determined by the dominant order parameter susceptibility $\bar{\chi}_{\psi}$ defined by
\begin{equation}
    \bar{\chi}_{\psi}=\frac{1}{N}\sum_{ij}\sum_{\alpha\beta}n_i^\alpha \bar{\chi}_{ij}^{\alpha\beta}n_j^\beta, 
    \label{eq:order-specific}
\end{equation}
where the vectors $\bm{n}_i$ describe fixed spin orientations on all sites $i$ corresponding to the classical magnetic order $\psi$ being probed~\cite{our_HDM_paper}. We take the orders $\psi$ from our irrep analysis, i.e., they are of $E$, $T_1^{xy}$, $T_1^{\rm Ice}$, $A_2$ or $T_2$ type. Note that the order parameter susceptibilities corresponding to different $\bm{q}=\bm{0}$ spin configurations within the same irrep are identical~\cite{our_HDM_paper}.
Note that in \eqref{eq:order-specific}, $\bar{\chi}^{\alpha\beta}_{ij} \equiv\int_0^\infty d\tau \langle S_i^\alpha(\tau) S_j^\beta(0)\rangle$ is the static spin correlation function computed in imaginary time domain $\tau$.

The quantum phase diagram in Fig.~\ref{fig:Q_HTSE+PFFRG}(a) determined using PFFRG shows distinct differences from the classical one in Fig.~\ref{fig:fig_1_structure_factors}(a):
 (i) The paramagnetic (gray) domain extends its regime of stability, expanding into the region of classical $T_1^{xy}$ order. (ii) The extent of the $E$ long-range ordered phase is enhanced by quantum fluctuations and partially penetrates the paramagnetic region of the classical model, in agreement with a previous quantum Monte Carlo study considering $J_{\pm\pm}=0$~\cite{KatoOnodaPRL2015}. (iii) These phase boundary shifts caused by quantum fluctuations create a characteristic narrow corridor of quantum paramagnetic behaviour approximately parallel to the classical $E$-$T_1^{xy}$ phase boundary.

The distinction between magnetic order and quantum paramagnetic behaviour via a flow breakdown of susceptibilities is subject to uncertainties in PFFRG especially along the phase boundaries. As shown in Fig.~\ref{fig:Q_HTSE+PFFRG}(a), the quantum DQQ model lies in such a region of uncertainty. 
Although the unambiguous identification of magnetic order with PFFRG is not possible in this region, we observe a comparatively large $\bar{\chi}_E$ order parameter susceptibility~\cite{SuppMat} indicating strong $E$-type spin correlations in the quantum DQQ model at $T=0$.
This is in stark contrast to the classical DQQ model which shows dominant spin-ice correlations in the low temperature limit.
However, we find that for coupling parameters deeper in the quantum paramagnetic corridor, but still close to the DQQ point, some of our observations from the classical DQQ model are recovered. Specifically, by visual inspection, we observe  matching spin structure factor patterns of the quantum non-Kramers model around $\{J_{zz},J_{\pm},J_{\pm\pm}\} = \{3.0,0.3,-1.0\}$ for $T=0$ and that of the classical DQQ model at $T\gtrsim T^*$, compare Fig.~\ref{fig:Q_HTSE+PFFRG}(c) with Fig.~\ref{fig:fig_1_structure_factors}(c) and Fig.~\ref{fig:Q_HTSE+PFFRG}(d) with Fig.~\ref{fig:fig_1_structure_factors}(d). 
The fact that this agreement holds only when the classical model is considered in the {\it intermediate} ($T^\ast \lesssim T \lesssim 
T_{\rm gl}$) temperature regime where the $R_1$-$R_2$ phase is realized suggests that {\it no} corresponding entropic selection mechanism occurs in the quantum model at $\{J_{zz},J_{\pm},J_{\pm\pm}\} = \{3.0,0.3,-1.0\}$ when the temperature is lowered towards $T\rightarrow 0$ as is observed in the classical model at the DQQ point. This is further confirmed by the renormalization group flows of order parameter susceptibilities $\bar{\chi}_{\psi}$, shown in Fig.~\ref{fig:Q_HTSE+PFFRG}(b), where $\bar{\chi}_{E}$, $\bar{\chi}_{T_1^{xy}}$, and $\chi_{T_1^{\rm Ice}}$ are all of similar size for $\{J_{zz},J_{\pm},J_{\pm\pm}\} = \{3.0,0.3,-1.0\}$ at $T=0$.
Note, however, that $\chi_{T_1^{\rm Ice}}$ only takes into account $\bm q = \bm 0$ spin ice configurations and thus does not represent the full spin ice manifold. All these observations lend support to the conclusion that in this corridor with paramagnetic behaviour, an exotic quantum spin liquid described by coexisting emergent rank-1 and rank-2 gauge fields may exist. 
On the other hand, it is strongly believed that the XXZ model in the $0 <J_{\pm} \lesssim 0.156$ window on the $J_{\pm\pm}=0$ axis hosts a U(1) spin liquid~\cite{KatoOnodaPRL2015}.
 It is likely that this phase somewhat extends over a finite 
$|J_{\pm\pm}| \neq 0$ range.
Thus, our results beg the question whether there is a quantum spin liquid to quantum spin liquid transition as the DQQ point is approached from above (i.e. from the $J_{\pm\pm}=0$ axis and $J_\pm$ becoming negative in Fig.~\ref{fig:Q_HTSE+PFFRG}(a) (and correspondingly for the DQQ$^*$ point; not shown).

To corroborate these observations with a complementary method, we next discuss the results of a high-temperature series expansion (HTSE) study, whose details can be found in Ref.~\cite{SuppMat}. We calculate the susceptibilities for both $T^{xy}_1$ and $E$ order-parameters, focusing on the coupling regime where PFFRG identified a paramagnetic corridor. The first order HTSE for the inverse susceptibility gives the Curie-Weiss law (taking the
zeroth order term as the single-spin Curie term). 
We find that the Curie-Weiss temperature for the $T_1^{xy}$ and $E$ 
 order parameters exchange dominance upon crossing the classical 
 $T_1^{xy}-E$  phase boundary. This shows the equivalence of classical and quantum models at the Curie-Weiss (mean-field) level.

Higher orders of the expansion are analyzed by Pad\'e approximants~\cite{oitmaa_book} and show the following: The $E$ susceptibility dominates along the classical $E$-$T_1^{xy}$ phase boundary and, to some degree, in the classical $T^{xy}_1$ phase. The blue circles in Fig.~\ref{fig:Q_HTSE+PFFRG}(a) mark regions where the $E$ susceptibility grows rapidly upon decreasing temperature, while the orange circles show regions where the $T^{xy}_1$ susceptibility grows rapidly. The parameter regions where the $T^{xy}_1$ and $E$ susceptibilities dominate, respectively, agree well with the  PFFRG results. 
In between these $T^{xy}_1$ and $E$ regions, HTSE finds a corridor running parallel to the classical $E$-$T_1^{xy}$ phase boundary, but shifted inside the classical $T^{xy}_1$ phase where neither susceptibility is found to grow substantially, again very similar to the paramagnetic strip found in PFFRG. Altogether, these results imply the exciting possibility of an extended region of a quantum spin liquid phase near the DQQ point, and  potentially exhibiting an unusual gauge structure.
\\
\section*{\bf Discussion} 
The prevalence of different gauge theories on different energy scales is a well-known feature of the Standard Model of High-Energy Physics, which comprises U(1), SU(2) and SU(3) gauge structures that produce distinct physical properties, e.g., on typical atomic and nuclear energy scales. In this paper, we have used the concept of coexisting gauge theories associated with different energy scales in a condensed matter system.
More precisely, the system we investigated is a Heisenberg plus Dzyaloshinskii-Moriya interaction model on the pyrochlore lattice in the vicinity of the classical multi-phase triple point $D/J=-2$. At the classical level, the multi-phase point has been studied by a comprehensive set of numerical and analytical methods and found to exhibit several unique features. There are two distinct spin liquid phases as a function of temperature 
with no signal of a symmetry-breaking transition down to the lowest temperatures. The two spin liquid phases uncovered are
described by different effective low-energy gauge theories with a sharp crossover between the two that is signaled by a peak in the specific heat. The higher temperature spin liquid phase is a novel $R_1$-$R_2$ spin liquid that exhibits both two-fold and four-fold pinch points in the spin structure factors and is described by fluctuating rank-1 vector and rank-2 tensor fields.

The lower temperature spin liquid, which persists down to $T=0$, is a spin-ice-like Coulomb phase whose emergence from the higher temperature phase is a novel illustration of entropic `disorder-by-disorder' selection. For the manifold of states selected at low temperatures, a fraction of the eigenmodes have zero energy at the quadratic level. This gives them an entropic advantage and is also responsible for reducing the low-temperature heat capacity below the standard equipartition value assuming solely quadratic-level spin fluctuations.

Our study of the quantum model using pseudo-fermion functional renormalization group (PFFRG) shows that phase boundaries are renormalized by quantum fluctuations, with the best match between the classical and quantum spin-spin correlations realized slightly away from the $D/J=-2$ value. Perhaps more interestingly, we find an extended corridor in parameter space that runs parallel to the classical phase boundary between the $T_1^{xy}$ and $E$ phases, where the ground state remains non-magnetic. These results are further supported by high-temperature series expansion calculations for different order-parameter susceptibilities.

The full anisotropic features of the spin-spin correlations are needed to distinguish between the two spin-liquid phases. The use of isotropic g-factors was essential to expose this difference in a compact manner. In non-Kramers spin systems, coupling to time-reversal odd fields such as external magnetic-fields or neutron-spin requires $g^{zz}\ne 0$, 
$g^{xx} = g^{yy}\equiv g^\perp = 0$~\cite{Lee_kramers_PhysRevB.86.104412,Rau-2019}. Structure factors obtained with these anisotropic $g$-factors fail to distinguish between the two phases (See Supporting Information). This means that experimental elucidation of the two spin-liquids and the transition between them would pose new challenges, but  which could possibly be tackled by studying response to strain-fields \cite{IanFisher}.

From a materials perspective, a previous work~\cite{TakatsuPRL2016TbtiO} suggested that the perplexing Tb$_2$Ti$_2$O$_7$ pyrochlore antiferromagnet~\cite{Gardner-Tb2Ti2O7,Rau-2019}, or its Tb$_{2+x}$Ti$_{2-x}$O$_7$ off-stoichiometry variant, may be located in the vicinity of the DQQ$^\ast$ point we identified in this work.
Our results provide an enlarged and intriguing perspective as to the exotic physics at play in these compounds; namely that they may reside in a region of spin-spin coupling parameters near the DQQ$^\ast$ point, dual to the DQQ point where quantum spin-liquidity is observed in Fig.~\ref{fig:Q_HTSE+PFFRG}(a).

 Our investigation raises several questions that should be addressed in the future: (i) Is the low-temperature spin-ice phase near the DQQ point continuously connected to the U(1) quantum spin ice~\cite{Gingras_2014} spin liquid around the Ising model? If not, is it possible to observe a transition by continuously tuning the interaction couplings as was predicted for other similar systems~\cite {yan2023experimentally,sanders2023vison}. (ii) Are there emergent U(1) photons in this phase and what signatures do they have?  (iii) How could one experimentally investigate the dynamical pseudospin structure factors associated with non-dipolar degrees of freedom?

Finally, the results presented in this paper are a further reminder of the richness of collective phenomena that frustrated spin systems harbour,  creating ``their own'' nanoscale structure of effective degrees of freedom and driving distinct thermal and quantum regimes of spin liquidity. 
Intriguingly, this is akin to the local tetrahedral bonding arrangement that some compounds, such as phosphorous, sulphur and silicon possess in their liquid state and which is thought to play an essential role in the liquid-to-liquid transition they display~\cite{S-L2L,P-L2L}.

\section*{Methods}

\noindent{\bf Irreducible representation analysis. }
As was noted in Ref.~\cite{Yan-2017}, the most general nearest-neighbour bilinear pyrochlore Hamiltonian can be written as the sum of single up and down tetrahedra Hamiltonians
\begin{eqnarray}
    \mathcal{H}=\sum_\boxtimes \mathcal{H}^\boxtimes,
\end{eqnarray}
where $\mathcal{H}^\boxtimes$ can be decomposed into its irreducible representations (irreps)~\cite{SuppMat}, see \eqref{eq:single_tet_Hamiltonian}, where the weights $a_I$ are smooth linear functions of the nearest-neighbour interaction couplings $\{J_{zz},J_{\pm}, J_{\pm\pm}, J_{z\pm}\}$~\cite{wong13,Yan-2017} and $\bm m_I^\boxtimes$ are the corresponding irrep fields onto which the spin configuration for tetrahedron $\boxtimes$ can be decomposed. In general, the irrep decomposition of the single tetrahedron Hamiltonian allows a coupling term between the $T_1^{xy}$ and the $T_1^{\rm Ice}$ irreps. This term, however, is proportional to the nearest-neighbour coupling $J_{z\pm}$ which vanishes for non-Kramers systems. We refer the reader to Refs.~\cite{our_HDM_paper,Yan-2017} for a more detailed discussion of the irrep fields $\{\bm m_I\}$ and how these are expressed in terms of the spins on a single tetrahedron. The irrep analysis can be used as a first approximation in the prediction of a long-range ordered phase~\cite{Yan-2017,our_HDM_paper}, as well as the building blocks in the construction of an effective long-wavelength theory for systems that avoid long-range order~\cite{SuppMat}.

To study the evolution of the spin configurations in our MC simulations as a function of temperature, we decomposed every single up-tetrahedron of an MC-sampled system into its irrep field and study the distribution of the magnitude of these fields as well as its average value for a set of temperatures, see Fig.~\ref{fig:fig_2_statistics_MC}. 
\\

\noindent{\bf Monte Carlo simulations.} Monte Carlo (MC) simulations were performed on systems of size $L\in\{6,8,10,12\}$, corresponding to $4L^3$ classical spins with $|\bm S_i|=1$, where we used $5\times 10^4$ thermalization sweeps and $8\times 10^4$ measurement sweeps. For each sweep, the system was updated using a Gaussian update~\cite{Alzate-Cardona_2019}, over-relaxation~\cite{ZhitomirskyPRL2012,CreutzPRD}, and a loop algorithm for Heisenberg spins which was inspired by Refs.~\cite{Shinaoka_2011,ShinaokaPhysRevB.82.134420}, where a single loop is attempted to be both identified and flipped per sweep. Additionally, we performed an average over $500$ independent MC simulations.

Additionally, we implemented a classical Ising MC with single spin-flip updates supplemented with a loop algorithm to sample the perfect spin-ice configurations of system size $L\in \{2,3,4,6,8,10,12\}$ used in the analysis of the classical low-temperature expansion.
\\

\noindent{\bf Self-consistent Gaussian approximation. }
The self-consistent Gaussian approximation (SCGA)~\cite{ConlonAbsentPhysRevB.81.224413,ChungKristianPhysRevLett.128.107201} is a classical approximation where the spin-length constraint is replaced by a soft-spin constraint where the spin length is preserved on \textit{average} over the whole system. In other words, for an $n$-component spin, the spin-length constraint, $\sum_{\alpha=1}^n\left(S^{\alpha}  \right)^2=|\bm S|^2=S^2$, is replaced by the thermodynamic average condition $\langle |\bm S|^2 \rangle =S^2$. To compare our results with the MC simulations we take $S=1$. This condition is enforced by the introduction of a Lagrange multiplier $\lambda$ imposing the average constraint at all temperatures. The introduction of this approximation in the spin-length constraint results in a quadratic (Gaussian) theory  which can be solved numerically exactly, and from which quantities such as the spin-spin correlation functions can be computed~\cite{SuppMat}. 
\\

\noindent{\bf Classical low-temperature expansion.}
A classical low-temperature expansion is a framework where a low-energy Hamiltonian describing the fluctuations about a low-temperature state is derived. Such Hamiltonian can be obtained for the general bilinear spin Hamiltonian~\cite{SuppMat} by assuming that the spin components are described in a local orthonormal frame where the (local) $\tilde{z}_i$  axis is along the zero-temperature orientation of the spin ${\bm S}_i$ at a given pyrochlore lattice site $i$. This allows us to write the spin at FCC site $i$  and sublattice $a$ as,
\begin{equation}
    \bm{S}_{ia}\simeq \left(\delta n^{\tilde x}_{ia},\delta n^{\tilde y}_{ia}, S\left(1-\frac{(\delta n^{\tilde x}_{ia})^2}{2S^2}-\frac{(\delta n^{\tilde y}_{ia})^2}{2S^2}\right)\right)\label{eq:fluctuations_CLTE},
\end{equation}
where it is assumed that the system displays small fluctuations $\delta n^\alpha_{ia}$ about the low-temperature spin orientation, where $\alpha$ labels the perpendicular directions to the low-temperature spin orientation with $|\delta n^\alpha | \ll S$, where we take $S=1$. Introducing the above expression into the Hamiltonian in \eqref{eq:CH6-H_H+DM}, and considering terms up to second  order in 
$\{\delta n^\alpha_{ia}\}$, results in a quadratic Hamiltonian
that can be used to determine the energy dependence of the fluctuations about a given ground-state configuration. 
More details regarding the implementation of this method are presented in the Supporting Information~\cite{SuppMat}.
\\

\noindent{\bf Pseudo-fermion functional renormalization group (PFFRG).}
In PFFRG, the spins are first mapped onto pseudo-fermions which allows one to study quantum spin models at $T=0$ within the functional renormalization group formalism~\cite{Reuther10,Mueller23}. After introducing an infrared frequency cutoff parameter $\Lambda$ in the fermionic propagator, coupled differential equations for the fermionic vertex functions are solved from the known high-energy limit $\Lambda\rightarrow\infty$ towards the cutoff-free model $\Lambda\rightarrow 0$. The computed $\Lambda$-dependent susceptibility reveals whether a model is magnetically ordered or quantum paramagnetic. Because of the approximations involved, a magnetic order transition usually does not result in a divergence of the susceptibility, but rather in a flow breakdown manifested by a kink. In contrast, a quantum paramagnetic susceptibility flow remains smooth down to the cutoff-free limit $\Lambda\rightarrow0$.

We apply the one-loop plus Katanin PFFRG method~\cite{Reuther10,Mueller23} with an exponential frequency mesh, containing $32^{3}$ ($1000$) positive frequencies for the fermionic two-particle vertex (self-energy). Spin correlations, spanning beyond a distance of four nearest-neighbour distances, are approximated to be zero (no periodic boundary conditions are applied). The flow equations are solved using an explicit embedded Runge-Kutta (2, 3) method with adaptive step size~\cite{GSL09}.
\\

\noindent{\bf High-temperature series expansion.} The high-temperature series expansion method is based on expanding the Boltzmann weight $\exp\left({-\beta {\cal H}}\right)$ in powers of the inverse temperature $\beta\equiv 1/T$, 
\begin{equation}
    \exp{\left(-\beta {\cal H}\right)} = \sum_n \frac{ (-\beta {\cal H})^n}{n!}.
\end{equation}
High-temperature series expansion for an extensive property $P$ can be calculated by a linked-cluster method \cite{oitmaa_book}, 
\begin{equation}
    \frac{P}{N_s}=\sum_c L(c) \times W(c),
    \label{Lattice-Constant}
\end{equation}
where $N_s$ is the number of spins, and the sum is over all linked or connected clusters that can be embedded in the lattice. The quantity $L(c)$, called the lattice constant, is the number of ways the cluster $c$ can be embedded in the lattice per lattice site. The quantity $W(c)$, called the weight of the cluster, can be obtained from a high-temperature series expansion for some physical 
property $P$ for the cluster $c$, $P(c)$, from the relation
\begin{equation}
    P(c) = \sum_{s\subseteq c} W(s),
\end{equation}
where the sum is over all subclusters of the cluster $c$ including the cluster $c$. Thus, starting with the smallest cluster, one can calculate the weight of all clusters up to some order. One can show that if all clusters of up to $n$ bonds are included in the sum in \eqref{Lattice-Constant} it gives 
the high-temperature series expansion for the infinite system to order $\beta^n$.

We have used the high-temperature series expansion method to calculate order-parameter susceptibilities for $T^{xy}_1$ and $E$ order parameters~\cite{SuppMat}.
\\



\begin{acknowledgments}
We acknowledge useful discussions with Kai Chung, Alex Hickey and Peter Holdsworth. 
The work at the University of Waterloo was supported by the NSERC of Canada and the Canada Research Chair (Tier 1, M.J.P.G.) program. 
Numerical simulations done at Waterloo were performed thanks to the computational resources of the Digital Research Alliance of Canada. DLG acknowledges the computing time provided by the Digital Research Alliance of Canada, and the financial support from the DFG through the Hallwachs-R\"ontgen Postdoc Program of the W\"urzburg-Dresden Cluster of Excellence on Complexity and Topology in Quantum Matter -- \textit{ct.qmat} (EXC 2147, project-id 390858490) and through SFB 1143 (project-id 247310070).
J.R. and V.N gratefully acknowledge the computing time provided to them on the high-performance computer Noctua 2 at the NHR Center PC2. 
This is funded by the Federal Ministry of Education and Research and the state governments participating on the basis of the resolutions of the GWK for the national high-performance computing at universities 
(\url{www.nhr-verein.de/unsere-partner}). 
Some of the computations for this research were performed using computing resources under project hpc-prf-pm2frg.
V.N. would like to thank the HPC Service of ZEDAT and Tron cluster service at the Department of Physics, Freie Universität Berlin, for computing time. 
The work of Y.I. and M.J.P.G. was performed, in part, at the Aspen Center for Physics, which is supported by National Science Foundation grant PHY-2210452. 
The work of R. R. P. S. was supported by the National Science Foundation grant DMR-1855111. J. O. acknowledges computing support provided by the Australian National Computation Infrastructure (NCI) program. J. R. thanks IIT Madras for a Visiting Faculty Fellow position under the IoE program. The participation of Y.I. at the Aspen Center for Physics was supported by the Simons Foundation. This research was supported in part by the National Science Foundation under Grant No.~NSF~PHY-1748958. Y.I. acknowledges support by the ICTP through the Associates Programme and from the Simons Foundation through grant number 284558FY19, IIT Madras through the QuCenDiEM CoE (Project No. SP22231244CPETWOQCDHOC), the International Centre for Theoretical Sciences (ICTS), Bengaluru, India during a visit for participating in the program “Frustrated Metals and Insulators” (Code: ICTS/frumi2022/9). Y.I. acknowledges the use of the computing resources at HPCE, IIT Madras.
\end{acknowledgments}
\bibliography{references}


\clearpage

\addtolength{\oddsidemargin}{-0.75in}
\addtolength{\evensidemargin}{-0.75in}
\addtolength{\topmargin}{-0.725in}

\newcommand{\addpage}[1] {
\begin{figure*}
  \includegraphics[width=8.5in,page=#1]{Supplementary_Material.pdf}
\end{figure*}
}

\addpage{1}
\addpage{2}
\addpage{3}
\addpage{4}
\addpage{5}
\addpage{6}
\addpage{7}
\addpage{8}
\addpage{9}
\addpage{10}
\addpage{11}
\addpage{12}
\addpage{13}
\addpage{14}
\addpage{15}
\addpage{16}
\addpage{17}

\end{document}